# Tunable Fractional Chern Insulators in Rhombohedral Graphene Superlattices

Jian Xie[1], Zihao Huo[1], Xin Lu[2], Zuo Feng[3], Zaizhe Zhang[1], Wenxuan Wang[1], Qiu Yang[1], Kenji Watanabe[4], Takashi Taniguchi[5], Kaihui Liu[3], Zhida Song[1,6], X.C. Xie[1,7,8], Jianpeng Liu[2]*, and Xiaobo Lu[1,6]*

[1]International Center for Quantum Materials, School of Physics, Peking University, Beijing 100871, China
[2]School of Physical Science and Technology, ShanghaiTech Laboratory for Topological Physics, ShanghaiTech University, Shanghai 201210, China
[3]State Key Laboratory for Mesoscopic Physics, Frontiers Science Centre for Nano-optoelectronics, School of Physics, Peking University, Beijing 100871, China
[4]Research Center for Electronic and Optical Materials, National Institute of Material Sciences, 1-1 Namiki, Tsukuba 305-0044, Japan
[5]Research Center for Materials Nanoarchitectonics, National Institute of Material Sciences, 1-1 Namiki, Tsukuba 305-0044, Japan
[6]Collaborative Innovation Center of Quantum Matter, Beijing 100871, China
[7]Institute for Nanoelectronic Devices and Quantum Computing, Fudan University, Shanghai 200433, China
[8]Hefei National Laboratory, Hefei 230088, China

*E-mail: xiaobolu@pku.edu.cn; liujp@shanghaitech.edu.cn;

**Fractional Chern insulators (FCIs) showing a transport effect with fractionally quantized Hall plateaus emerging under zero magnetic field, provide a radically new opportunity to engineer topological quantum electronics[1-3]. By construction of topological flat band with moiré engineering, intrinsic FCIs have been observed in twisted $MoTe_2$ system[4-7] and rhombohedral pentalayer graphene/hBN moiré superlattices[8] with anomalous Hall resistivity quantization number $C \leq 2/3$ including the gapless composite Fermi-liquid state with $C = 1/2$. Here, we experimentally demonstrate a new system of rhombohedral hexalayer graphene (RHG)/hBN moiré superlattices, which exhibit both integer and fractional quantum anomalous Hall effects with rich tunability including electric displacement field, perpendicular magnetic field and in-plane magnetic field. By tuning the electrical and magnetic fields at $0 < \nu < 1$, we have observed a quantum phase transition showing a sign reversal of the Hall resistivity at finite magnetic fields. Surprisingly, the FCI state at $\nu = 2/3$ survives in the phase transitions, exhibiting a robust quantized Hall resistivity across both phases. Finally we have further demonstrated the indispensable role moiré potential plays in the formation of the flat Chern band from a theoretical perspective. Our work has established RHG/hBN moiré superlattices as a promising platform for exploring quasi-particles with fractional charge and non-Abelian anyons[9-13] at zero magnetic field.**

Fractional Chern insulators (FCIs), exhibiting the fractional quantum anomalous Hall effect (FQAHE)[1-3], exemplify exotic quantum phenomena in topological flat bands where interactions dominate beyond single-particle physics. The fractionally charged quasiparticle excitations in fractional quantum anomalous Hall insulators provide new opportunities to realize topological quantum computation[13] at zero magnetic field. Recently, in several pioneering works, FQAHE has been experimentally observed in the flat moiré bands of twisted $MoTe_2$ system[4-7] and rhombohedral pentalayer graphene/hBN moiré superlattices[8]. Fundamentally, FQAHE is a lattice generalization of the fractional quantum Hall effect[14-21] (FQHE) where hundreds of fractionally quantized states have been observed, encompassing the vast majority of the quantized plateaus. Experimentally, FQAHE is challenging to be achieved as it requires synergism from many factors including strong electron-electron interactions, isolated flat band, uniform Berry curvature distribution and desirable quantum geometric properties[1-3,22-25]. Moiré superlattices of two-dimensional materials offer new opportunities to optimize these quantum effects in several aspects: the interacting strength can be tuned by controlling the electrical displacement fields and twist angles; the uniformly distributed Berry curvature can be more readily realized in a Brillouin zone with a significantly reduced size due to band folding induced by moiré potential (typical wavelength $\lambda \sim 10$ nm); and the carrier density $n_0 \sim 10^{12}$ cm$^{-2}$ corresponding to the full filling of a flat Chern band analogous to the lowest Landau level, can be continuously supplied by electrical gating. Motivated by several experimental and theoretical works on rhombohedral multilayer graphene[8,26-40] showing flat band at low Fermi energy and particularly on rhombohedral pentalayer graphene exhibiting an abundance of FCIs, in this work we investigate the rhombohedral hexalayer graphene (RHG)/hBN moiré superlattices.

**Topological flat band in RHG moiré superlattices**

Figure 1a illustrates details of device 1 in which the RHG is encapsulated by hBN layers and further aligned with the bottom-layer hBN at a twist angle $\theta \approx 0.17°$ (moiré periodicity $\lambda \approx 14.5$ nm). The carrier density $n$ and displacement field $D$ can be subsequently tuned by top and bottom graphite gates (Methods). Figure 1b shows the longitudinal resistivity $\rho_{xx}$ as a function of displacement field $D$ and moiré filling factor $v = n/n_0$ measured at zero magnetic field. Similar to previously reported rhombohedral graphene moiré superlattices[8,26,27], the phase diagrams exhibit rich correlated phenomena, i.e., the correlated insulators at $v = \pm 1$ and $\pm 2$. However, the layer antiferromagnetic (LAF) state at $D/\varepsilon_0 = 0$ and $v = 0$ observed in other systems[8,30,31] has not been observed in device 1. One possibility is that the LAF state in RHG can be suppressed by moiré modulations especially when the twist angle is close to zero.

Figure 1c displays detailed measurements of $\rho_{xx}$ and $\rho_{xy}$ in the region indicated by the yellow dashed box in Fig. 1b. The $v = 1$ state corresponding to a fully filled flat conduction band with spontaneously split SU(4) isospin (valley-spin) flavor, starts to emerge at $D/\varepsilon_0 \approx -0.45$ V nm$^{-1}$. In the meanwhile, $\rho_{xx}$ immediately vanishes and $\rho_{xy}$ develops finite values with increasing $D$. The behaviors of $\rho_{xx}$ and $\rho_{xy}$ at $v = 1$, particularly the nearly perfect quantization of $\rho_{xy} = h/e^2$, highly indicate the formation of an intrinsic Chern insulator with $|C| = 1$. Detailed characterization of the $v = 1$ at fixed $D$ and sweeping $B_\perp$ fields reveals the integer quantum anomalous Hall effect (IQAHE) evidenced by a standard magnetic hysteresis loop with $\rho_{xy}$ quantized at $h/e^2$ (with an accuracy > 99.5%) and vanishing $\rho_{xx}$. At the coercive field $B_c \approx \pm 12$ mT, the longitudinal resistivity $\rho_{xx}$ remains on a vanishing level, and Hall resistivity abruptly changes its sign, corresponding to a transition between states with opposite magnetizations. Similar behaviors have been observed in a second device in which RHG is also aligned with bottom hBN at a twist angle $\theta \approx 0.22°$ (moiré periodicity $\lambda \approx 14.3$ nm) shown in Extended Data Fig. 8a-d. Moreover, different from Chern insulators reported in other moiré systems[8,41], the $|C| = 1$ state here can be stabilized at zero magnetic field even when tuning the carrier density as shown in Fig. 1f, implying a spontaneous time-reversal breaking ground state with large magnetic anisotropy. Note that the quantized $\rho_{xy}$ shown in Fig. 1f and g is without any symmetrized or antisymmetrized processing (Extended Data Fig. 3), and can be sustained up to a temperature of $T = 1.5$ K (Extended Data Fig. 5a). Similar to the pentalayer graphene system[8], the $|C| = 1$ state in RHG also appears in one side of the phase diagram (Fig. 1b) where the electrons are pushed away from the moiré interface under $D$ field. Our experimental results can qualitatively match with the interacting band structure calculated using an approach combining renormalization-group, constrained random-phase-approximation (cRPA), and Hartree-Fock methods (Fig. 1d and e). Moreover, we note that at large negative displacement fields ($D/\varepsilon_0 < -0.6$ V nm$^{-1}$), the values of $\rho_{xy}$ also exceed $h/e^2$ over a wide range of moiré fillings at $1/2 < \nu < 1$, highly implying the formation of zero-field FCIs.

**The FCI at ν = 2/3**

The zoomed-in phase diagram of $\rho_{xx}$ versus $D$ and $\nu$ measured in device 1 (Fig. 2a), shows a clear dip (marked with the red dashed line) near $\nu = 2/3$. Further measurement of Fig. 2b displays the symmetrized $\rho_{xx}$ and antisymmetrized $\rho_{xy}$ obtained at $B_\perp = \pm 40$ mT and $\nu = 2/3$ as a function of $D$ fields. A continuous phase transition from Fermi liquid state at low $|D|$ fields to FCI state at high $|D|$ fields has been observed. At an optimal $D$ field around $D/\varepsilon_0 = -0.75$ V nm$^{-1}$, the $\rho_{xy}$ reaches a maximum value of $3h/2e^2$ (with an accuracy > 99%) and the $\rho_{xx}$ reaches a minimum. Figure 2c shows the symmetrized $\rho_{xx}$ and antisymmetrized $\rho_{xy}$ as a function of $\nu$ measured at the optimal $D$ field. Consistent with Fig. 2b, the $\rho_{xy}$ is quantized to $3h/2e^2$ (with an accuracy > 99%) at $\nu = 2/3$ and $\rho_{xx}$ shows a local minimum, indicating an incompressible FCI state with $|C| = 2/3$. Magnetic hysteresis loops are shown in Fig. 2d and e, exhibiting FQAHE which can survive in a temperature up to 400 mK (Extended Data Fig. 5d). Similar to $\nu = 1$ state, we have not observed the resistance peaks of $\rho_{xx}$ in the magnetic loops which are usually accompanied by the magnetization flips, even the sweeping steps of $B_\perp$ field are down to 0.5 mT. Generally, the resistance peaks of $\rho_{xx}$ in the magnetic loop originate from the breakdown of the non-dissipative FQAHE/IQAHE due to fluctuations in magnetization near the coercive fields. The absence of resistance peaks of $\rho_{xx}$ indicates first-order-like magnetization flips due to the large magnetic anisotropy which is consistent with the orbital nature of magnetization. Strikingly, with finite $B_\parallel$ fields applied, the magnetic loops show broader transition regions near the coercive field in $\rho_{xy}$ with peaks in $\rho_{xx}$ appearing simultaneously (Fig. 2d and e). The magnetic hysteresis loop is completely suppressed when $B_\parallel > 1.5$ T. The detailed mechanism behind the suppression of the FCI state under $B_\parallel$ fields is still a mystery particularly considering the weak spin-orbit coupling and the much larger magnetic length ($l_B = \sqrt{\hbar/eB} \approx 20$ nm when $B = 1.5$ T) compared to the thickness of hexalayer graphene (~ 2 nm). One possible explanation is to consider the effective out-of-plane momentum induced by the in-plane orbital effects of $B_\parallel$ field, which has also been adopted in twisted multilayer graphene systems[42]. The additionally gained out-of-plane momentum, which is proportional to $B_\parallel$ field can lead to reconstructions of Fermi surface and break the ground state of FCIs.

**Phase transitions in the $B_\perp$ field within $0 < \nu < 1$**

The $\rho_{xx}$ and $\rho_{xy}$ versus $\nu$ and $B_\perp$ field at a fixed displacement field of $D/\varepsilon_0 = -0.74$ V nm$^{-1}$ are shown in Fig. 3a and b. Notably, the phase diagram is divided into different regions. The tilted evolution of $\nu = 2/3$ gap at low $B_\perp$ fields shown in Fig. 3a can be well explained by Streda's formula $\frac{\partial n}{\partial |B_\perp|} = \frac{Ce}{h}$ in a transport regime. Whereas the evolution of $\nu = 2/3$ gap abruptly switches to an opposite tilted direction with increasing $B_\perp$ fields, but still agrees well with Streda's formula in terms of the magnitude of slope, indicating a sign reversal of Hall resistivity. This is further evidenced by $\rho_{xy}$ in Fig. 3b which exhibits multiple sign reversals. Figure 3c shows $\rho_{xy}$ versus $B_\perp$ field and $D$ field at $\nu = 2/3$. The evolution of four phases corresponding to different configurations of magnetizations and carrier types can be clearly resolved when the magnitude of $D$ field exceeds a critical value of $D_c/\varepsilon_0 = -0.62$ V nm$^{-1}$. Figure 3d and e exhibit $\rho_{xx}$ and $\rho_{xy}$ versus $B_\perp$ field extracted along the dashed lines from Fig. 3a, b and Extended Data Fig. 6. Strikingly, the $\nu = 2/3$ state of $C = -2/3$ in the higher $B_\perp$ fields persists with a quantized value of $\rho_{xy} \approx 3h/2e^2$. Similar behaviors have been observed in device 2 (Extended Data Fig. 8e-i) with signatures appearing at more fractional fillings including $\nu = 2/3, 3/5$ and $4/7$.

Generally, the sign of $\rho_{xy}$ depends on the direction of magnetization and sign of the integrated Berry curvature, which corresponds to the Chern number for a gapped state. As the $B_\perp$ field tends to favor parallelly aligned magnetization, changing the direction of $B_\perp$ field will flip the magnetization and lead to a sign reversal of $\rho_{xy}$ as shown in Fig. 3b at $B_\perp = 0$. The additional sign reversal of $\rho_{xy}$ at $B_\perp \approx 0.8$ T suggests a change in the sign of the integrated Berry curvature (Chern number for the $\nu = 2/3$ state). In graphene systems, changing either valley or carrier type results in a sign reversal of the integrated Berry curvature. As shown in Fig. 3f, we have proposed two possible mechanisms for the phase transition at finite magnetic fields. One possibility is that the system switches to the K' valley which hosts a Berry curvature opposite to that of the K valley, as illustrated in the left panel of Fig. 3f. Such transitions induced by valley polarization have been observed in several systems[44-49], where the reversible $\rho_{xy}$ typically exhibits hysteretic behavior when sweeping gate voltage or magnetic field back and forth. These behaviors can be attributed to the competition between the orbital magnetism originating from bulk states and edge states within the Chern gap[50]. However, this differs somewhat from the phase transition we observed here. Firstly, there is no hysteresis when sweeping electrical or magnetic field back and forth across the phase boundary at finite $B_\perp$ fields, suggesting a non-first-order phase transition. Secondly, unlike previously reported systems where the reversible $\rho_{xy}$ exists near a Chern gap, the phase transition shown in Fig. 3a and b occurs over a wide range of $\nu$ and is insensitive to the chemical potential. The other possibility is a reshuffle of charge carriers from electron type at low $B_\perp$ fields to hole type at high $B_\perp$ fields, where the two states correspond to the same valley polarization (right panel shown in Fig. 3f). In this scenario, the $\nu = 2/3$ state here shows a phase transition corresponding to a state hosting composite quasi-electron excitation (one electron dressed with two flux quanta) to the one hosting composite quasi-hole (one hole dressed with two flux quanta) excitation, which is in analogy to the fractional quantum Hall state at $\nu = 2/3$ described by the composite fermion picture[17,18]. Moreover, the magnetic flux in the composite fermions of the $\nu = 2/3$ FCI state arises from the intrinsic orbital magnetism due to interaction-driven spontaneous time-reversal symmetry breaking, which typically dominates over the field-induced magnetization of the Chern band. Notably, the FCI state with $C = -2/3$ at higher magnetic fields can survive at higher temperatures than that with quasi-electron excitations at lower magnetic fields (Fig. 3e and Extended Data Fig. 6). One possibility is that the magnetic field can further optimize the quantum geometry for stabilizing the FCI as shown in previous studies[43]. Alternatively, the magnetic field is responsible for stabilizing the magnetic orders, leading to more stable FCIs. The exact nature of the phase transition still needs to be further examined, whereas the coexistence of the FCI states at ν = 2/3 within both phases paves a unique avenue to tune FCIs and could experimentally illuminate the anti-FCI picture theoretically proposed recently[51].

**The role of moiré potential for the flat Chern band**

Similar to the pentalayer system[8], FCIs in RHG emerge under $D$ field which pushes away the electrons from the moiré interface. However, the threshold field ($|D/\varepsilon_0| \sim 0.7$ V nm$^{-1}$) in RHG is weaker compared to the pentalayer system ($|D/\varepsilon_0| \sim 0.9$ V nm$^{-1}$). To theoretically understand this, we have calculated the effective interaction strength of the systems versus $D$ field (Fig. 4a). The effective interaction strength of RHG under $|D/\varepsilon_0| \sim 0.6$ V nm$^{-1}$ is already comparable to that of the pentalayer system under $|D/\varepsilon_0| > 1$ V nm$^{-1}$. Moreover, the effective interaction strength of the RHG reaches a peak for 0.7 V nm$^{-1} \leq |D/\varepsilon_0| \leq 0.8$ V nm$^{-1}$, which precisely corresponds to the range of $D$ field within which the FCIs emerge.

Furthermore, we show that the moiré potential still plays an indispensable role in the formation of the flat Chern band although the flat Chern band is pushed away from moiré under large displacement field. We scale the moiré potential from 0% to 100% and study whether the Chern insulator state at ν = 1 exists at different moiré potential strengths using the cRPA + Hartree-Fock method. Compared with Hartree-Fock method, the cRPA+Hartree-Fock method we used here further takes into account the wavevector dependence of dielectric screening resulting from high-energy electron-hole fluctuations (Method). In Fig. 4c-h, we present the cRPA+Hartree-Fock band structures of RHG at ν = 1 for different moiré potential strengths: 0%, 20%, 40%, 60%, 80%, and 100%. We find that a gapped Chern insulator state emerges only at 100% moiré potential. For weaker moiré potential strengths, although the system does have non-vanishing valley and spin splitting, it still remains gapless at ν = 1. Moreover, we have calculated the real-space charge density distributions at 0% and 100% moiré-potential strengths as shown in Fig. 4b. Clearly, the charge density of the gapless Hartree-Fock state at 0% moiré potential strength is homogeneous everywhere, with a vanishing standard deviation. In contrast, the charge density of the Chern insulator state at 100% moiré potential exhibits clear charge modulation that is commensurate with the moiré superlattice. Our results clearly indicate that even at the cRPA+Hartree-Fock level, which neglects low-energy quantum fluctuation effects, moiré potential is indispensable for the Chern insulator state at ν = 1 in RHG system with a specific twist angle of ~ 0.2°. In the absence of moiré potential, the system remains in a gapless state with homogeneous charge density.

In conclusion, we have reported a new system composed of rhombohedral hexalayer graphene/hBN moiré superlattices, showing both integer and fractional Chern insulators. The FCI state at ν = 2/3 exhibits rich tunability with $D$ field, $B_\perp$ field and $B_\parallel$ field. The sign reversal of $\rho_{xy}$ at ν = 2/3 triggered by $B_\perp$ field phenomenologically suggests a quantum phase transition based on switching valleys or carrier types, whereas the robust FCI state at ν = 2/3 can survive at both phases. Given the high tunability of the system, further experiments with more tuning knobs, i.e., twist angles, dielectric environment and gate confinements, have great potential to shed light on the study of fractional charge and possible non-Abelian anyons.

We are grateful for fruitful discussions with A. Bernevig, L. Ju, D. Efetov, N. Regnault, J.R. Shi, K. T. Law, Q.L. He, L.D. Xian and A. Rubio. X.B.L. acknowledges support from the National Key R&D Program (Grant nos. 2022YFA1403500/02) and the National Natural Science Foundation of China (Grant Nos. 12274006 and 12141401). X.X. acknowledges support from Innovation Program for Quantum Science and Technology (Grant no. 2021ZD0302400). K.W. and T.T. acknowledge support from the JSPS KAKENHI (Grant nos. 21H05233 and 23H02052) and World Premier International Research Center Initiative (WPI), MEXT, Japan.

X.B.L. and J.X. conceived and designed the experiments; J.X. fabricated the devices and performed the transport measurement with help from Z.H., Z.F., Z.Z., W.W., Q.Y. and K.L.; J.X., Z.H., Z.S., X.X., J.L. and X.B.L. analyzed the data; X.L. and J.L. performed the theoretical modeling; T.T. and K.W. contributed hBN substrates; J.X., Z.H., J.L. and X.B.L. wrote the paper.

The authors declare no competing financial and non-financial interests.

## Methods
### Device fabrication

The RHG moiré samples are encapsulated with top and bottom hBN (~ 20 nm thick). Two graphite flakes (2~5 nm thick) are used to apply top and bottom gate voltages. All the flakes were firstly exfoliated onto $SiO_2$ (around 300 nm)/Si substrate, followed by detailed characterization with optical microscopes. The layer number of graphene flakes was identified via optical contrast which has been calibrated under background intensity (Extended Data Fig. 1). The rhombohedral region of hexalayer graphene was identified via the rapid infrared imaging technique[52] and further confirmed by Raman spectroscopy. The rhombohedral stacking domains of the hexalayer graphene were further delineated into isolated pieces through a femtosecond laser (pulse width ~ 150 fs, central wavelength ~ 517 nm, repetition frequency ~ 80 MHz and maximum power ~ 150 mW). Additionally, we meticulously selected straight edges of the flake with angles of $\pi/6$ or $\pi/3$ to each other as reference lines for alignment between bottom hBN and the RHG.

We initially picked up a hBN flake and a few-layer graphene flake by a polydimethylsiloxane (PDMS) stamp covered with a polypropylene carbonate (PPC) film and then released them onto a $SiO_2$/Si substrate to get the pre-fabricated bottom gate under RHG. As fabricated structure was then annealed in Ar/$H_2$ mixture atmosphere at 350 °C, followed by mechanical cleaning using the contact-mode AFM. For device 1, an auxiliary hBN flake (for picking up top graphite gate), top graphite gate, top hBN as dielectric layer for top gate and RHG were subsequently picked with a PPC/PDMS stamp and then dropped onto the pre-fabricated structure of bottom gate. For device 2, we used PC (poly bisphenol A carbonate)/PDMS stamp to subsequently pick up top graphite gate, top hBN as dielectric layer for top gate and RHG and then dropped onto the pre-fabricated structure of bottom gate. For the final stacks, we used Raman spectroscopy once more to verify that the sample regions maintain rhombohedral stacking. We utilized the technique of electron-beam lithography and reactive-ion etching to sculpt the device into Hall bar form. Finally, electrical connections were established by edge contact of Cr/Au (5 nm/70 nm) electrodes.

**Electrical transport measurement**

The device was measured with standard low-frequency lock-in techniques in a dilution refrigerator, with a base phonon temperature of approximately 10 mK. $R_{xx}$ and $R_{xy}$ were measured with Stanford Research Systems SR860 lock-in amplifiers combined with SR560 voltage preamplifier, with an excitation current of about 1 nA at a frequency of 17.777 Hz. Dual-gate voltages were applied using Keithley 2400 source-meters.

The carrier density $n$ and electrical displacement field $D$ were defined by top gate voltage $V_{tg}$ and bottom gate voltage $V_{bg}$, following $n = (C_{tg}V_{tg} + C_{bg}V_{bg})/e$ and $D = (C_{tg}V_{tg} - C_{bg}V_{bg})/2$, where $C_{tg}$ and $C_{bg}$ are the capacitances of the top and bottom gates per unit area, respectively, which were calibrated by the positions of different quantum Hall states (Extended Data Fig. 4 and 8).

**Determination of the alignment**

RHG was aligned with bottom hBN and misaligned with top hBN in our two samples. For device 1, the alignment between graphene and both hBN flakes can be easily identified with optical image. The well-defined zig-zag or armchair directions can be identified by finding two straight edges with exact 30° angle. Extended Data Fig. 2a shows the optical image of device 1 with the crystalline directions of different flakes indicated with different lines. The twist angle between graphene and top hBN is ~ 4° (corresponding to a moiré wavelength of ~ 3 nm) or ~ 26°. According to the optical image, the bottom hBN is well aligned with graphene which is consistent with the large moiré wavelength (~ 14.5 nm, corresponding to a twisted angle of 0.17°) obtained from transport measurements (Extended Data Fig. 4).

For device 2, the graphene could be aligned with both top and bottom hBN with twist angles of 0° or 30° according to the optical image (Extended Data Fig. 2b). Further measurement of second-harmonic generation (SHG) demonstrates a twist angle of 30° between top and bottom hBN, ruling out the possibly that the RHG is double aligned (0°) with both top and bottom hBN. Experimentally, there is a significant difference in the characteristic peaks of Raman spectrum between rhombohedral graphene samples with moiré and without moiré. Specifically, the 2D Raman peak of ABC graphene splits into two shoulders at wave numbers of 2700 cm$^{-1}$ and 2720 cm$^{-1}$. ABC graphene without notable moiré modulation exhibits a stronger peak at 2700 cm$^{-1}$ as shown in middle (RHG+top hBN in device 2 measured during stacking) and bottom panel (RHG+hBN with a large twist angle of ~ 15°) of Extended Data Fig. 2d. After the graphene was encapsulated with bottom hBN, the behavior of 2D peak of graphene changes significantly with the shoulder at 2700 cm$^{-1}$ notably suppressed (top panel of Extended Data Fig. 2d). The Raman characterization is also consistent with previous reported results[34]. With these systematic characterizations including optical image, SHG, Raman spectroscopy and transport measurements, the alignment of device 2 can be well identified. RHG in device 2 is ~ 30° aligned with top hBN and ~ 0.22° aligned with bottom hBN.

**Symmetrization and antisymmetrization**

Considering that $R_{xx}$ and $R_{xy}$ are symmetric and antisymmetric with respect to the magnetic field, respectively, we symmetrized/antisymmetrized them separately during the data processing, as follows:

$$R_{xx}^{Sym}(B,\nu) = \frac{R_{xx}^{Original}(B,\nu) + R_{xx}^{Original}(-B,\nu)}{2},$$

$$R_{xy}^{AntiSym}(B,\nu) = \frac{R_{xy}^{Original}(B,\nu) - R_{xy}^{Original}(-B,\nu)}{2}.$$

This approach helps eliminate the antisymmetric/symmetric parts in the $R_{xx}/R_{xy}$ data caused by other unfavorable factors such as deviation of standard Hall bar geometry. For the same reason, we obtained $R_{xx}$ and $R_{xy}$ data of the FQAHE at $B = 0$ from measurement data at opposite $B$,

$$R_{xy}(0) = \frac{R_{xy}(B) - R_{xy}(-B)}{2}, \quad R_{xx}(0) = \frac{R_{xx}(B) + R_{xx}(-B)}{2}$$

Note that here, the magnetic field should exceed the coercive field of the FQAHE.

Additionally, the similar procedures of symmetrization and antisymmetrization were employed to the magnetic hysteresis measurement data,

$$R_{xx,Forward}^{Sym}(B) = \frac{R_{xx,Forward}^{Original}(B) + R_{xx,Backward}^{Original}(-B)}{2},$$

$$R_{xx,Backward}^{Sym}(B) = R_{xx,Forward}^{Sym}(-B),$$

$$R_{xy,Forward}^{AntiSym}(B) = \frac{R_{xy,Forward}^{Original}(B) - R_{xy,Backward}^{Original}(-B)}{2},$$

$$R_{xy,Backward}^{AntiSym}(B) = -R_{xy,Forward}^{AntiSym}(-B).$$

And to more clearly demonstrate the features, resistivity $\rho$ is used instead of resistance $R$ for characterization

$$\rho_{xx} = R_{xx}\frac{w}{L}, \quad \rho_{xy} = R_{xy},$$

where the length-width ratio of the channel $w/L$ is 1/3 for device 1 and 1/2 for device 2.

**Theoretical calculation model**

In our theoretical study, we adopt the continuum model derived in previous work[53] to twisted hexalayer graphene-hBN moiré superlattices. The effective Hamiltonian describing the low-energy physics of hexalayer graphene includes two parts: the Hamiltonian of hexalayer graphene with rhombohedral stacking denoted as $H_{hexa}^{(0,\mu)}$, and the moiré potential $V_{hBN}(\mathbf{r})$ exerted on the bottom-layer graphene that is closest to the aligned hBN substrate (Supplementary Information). When *D* field is applied to the twisted RHG/hBN moiré system, it redistributes graphene's layer charge density, which in turn partially screens *D* field. Such a partial screening process can be treated by solving the classical Poisson equation in electrostatics-self consistently.

In graphene-based systems, the electrons in the filled high-energy bands (away from Dirac points) will act through long-ranged Coulomb potential upon the dynamics of low-energy electrons (near Dirac points), such that an effective low-energy Hamiltonian would have interaction-renormalized parameters different from the non-interacting ones. We take into account this effect using perturbative renormalization group approach, and the interaction-renormalized continuum model parameters are given in previous work[38] and Supplementary Information. In practice, one needs to define a low-energy cutoff $E_C^*$ to separate the low-energy and high-energy Hilbert space. Here $E_C^*$ is chosen in such a way that there are 3 valence and 3 conduction moiré bands per spin per valley within $E_C^*$.

We then consider *e-e* interactions between the low-energy electrons for $n_{\text{cut}} = 3$ moiré valence and conduction bands around charge neutrality point. Coulomb interactions projected to the Bloch wavefunctions of the low-energy bands are expressed as

$$\hat{V}^{\text{intra}} = \frac{1}{2N_s} \sum_{\tilde{\mathbf{k}},\tilde{\mathbf{k}}',\tilde{\mathbf{q}}} \sum_{\substack{\mu\mu' \\ \sigma\sigma' \\ ll'}} \sum_{\substack{nm \\ n'm'}} \left( \sum_{\mathbf{Q}} V_{ll'}^{\text{cRPA}}(\mathbf{Q}+\tilde{\mathbf{q}}) \, \Omega_{nm,n'm'}^{\mu l,\mu' l'}(\tilde{\mathbf{k}},\tilde{\mathbf{k}}',\tilde{\mathbf{q}},\mathbf{Q}) \right) \quad (1)$$
$$\times \hat{c}_{\sigma\mu,n}^\dagger(\tilde{\mathbf{k}}+\tilde{\mathbf{q}}) \hat{c}_{\sigma'\mu',n'}^\dagger(\tilde{\mathbf{k}}'-\tilde{\mathbf{q}}) \hat{c}_{\sigma'\mu',m'}(\tilde{\mathbf{k}}') \hat{c}_{\sigma\mu,m}(\tilde{\mathbf{k}})$$

where $\Omega_{nm,n'm'}^{\mu l,\mu' l'}(\tilde{\mathbf{k}},\tilde{\mathbf{k}}',\tilde{\mathbf{q}},\mathbf{Q})$ is the form factor due to the projection to the low-energy bands, and $V_{ll'}^{\text{cRPA}}(\mathbf{Q}+\tilde{\mathbf{q}})$ is the layer-dependent screened Coulomb interaction at wavevector $\mathbf{Q}+\tilde{\mathbf{q}}$. Here $\tilde{\mathbf{k}}, \tilde{\mathbf{k}}'$ and $\tilde{\mathbf{q}}$ are wavevectors within moiré Brillouin zone, and $\mathbf{Q}$ is moiré reciprocal vector. $\mu$, $\sigma$, and $l$ represent the valley, sublattice, and layer indices, respectively. More details are given in Supplementary Information.

The Coulomb interaction between two low-energy electrons can be screened due to the virtual excitations of particle-hole pairs formed outside the low-energy Hilbert space, which can be characterized using the constrained random phase approximation (cRPA). The cRPA has been successfully applied to twisted bilayer graphene[54,55]. Then, the cRPA screened Coulomb interaction is expressed as

$$\hat{V}_{ll'}^{\text{cRPA}}(\tilde{\mathbf{q}}) = \hat{V}_{ll'}(\tilde{\mathbf{q}}) \cdot \left(\hat{\epsilon}(\tilde{\mathbf{q}})\right)^{-1} \quad (2)$$

where $\hat{V}_{ll'}(\tilde{\mathbf{q}})$ is the original, layer-dependent Coulomb interaction expressed in matrix form, with $\hat{V}_{ll'}(\tilde{\mathbf{q}})_{\mathbf{Q},\mathbf{Q}'} = V_{ll'}(\tilde{\mathbf{q}}+\mathbf{Q})\delta_{\mathbf{Q},\mathbf{Q}'}$ ($V_{ll'}(\tilde{\mathbf{q}}+\mathbf{Q})$ is the Fourier transformed Coulomb interaction at wavevector $\tilde{\mathbf{q}}+\mathbf{Q}$). The cRPA dielectric constant is expressed as

$$\hat{\epsilon}_{\mathbf{Q},\mathbf{Q}'}(\tilde{\mathbf{q}}) = \left(1 + \chi^0(\tilde{\mathbf{q}}) \cdot \hat{V}(\tilde{\mathbf{q}})\right)_{\mathbf{Q},\mathbf{Q}'} \quad (3)$$

where $\chi^0(\widetilde{\mathbf{q}})$ is the susceptibility matrix at $\widetilde{\mathbf{q}}$ defined in the reciprocal-vector space. We then apply the Hartree-Fock approximation to Eq. (1) and solve for the ground state at filling $v = 1$.

**References**

1. Regnault, N. and Bernevig, B.A. Fractional Chern insulator. *Phys. Rev. X* 1, 021014 (2011).
2. Sheng, D. N., Gu, Z. C., Sun, K. & Sheng, L. Fractional quantum Hall effect in the absence of Landau levels. *Nat. Commun.* 2, 389 (2011).
3. Neupert, T., Santos, L., Chamon, C. and Mudry, C. Fractional quantum Hall states at zero magnetic field. *Phys. Rev. Lett.* 106, 236804 (2011).
4. Cai, J., Anderson, E., Wang, C. et al. Signatures of fractional quantum anomalous Hall states in twisted MoTe2. *Nature* 622, 63–68 (2023).
5. Zeng, Y., Xia, Z., Kang, K. et al. Thermodynamic evidence of fractional Chern insulator in moiré MoTe2. *Nature* 622, 69–73 (2023).
6. Park, H., Cai, J., Anderson, E. et al. Observation of fractionally quantized anomalous Hall effect. *Nature* 622, 74–79 (2023).
7. Xu, F., Sun, Z., Jia, T. et al. Observation of integer and fractional quantum anomalous Hall effects in twisted bilayer MoTe$_2$. *Phys. Rev. X.* 13, 031037 (2023).
8. Lu, Z., Han, T., Yao, Y. et al. Fractional quantum anomalous Hall effect in multilayer graphene. *Nature* 626, 759–764 (2024).
9. Arovas, D. and Schrieffer, J. R. and Wilczek, F. Fractional statistics and the quantum Hall Effect. *Phys. Rev. Lett.* 53, 722 (1984).
10. You, Y.Z. and Jian, C.M. and Wen, X.G. Synthetic non-Abelian statistics by Abelian anyon condensation. *Phys. Rev. B* 87, 045106 (2013).
11. Nakamura, J., Liang, S., Gardner, G.C. et al. Direct observation of anyonic braiding statistics. *Nat. Phys.* 16, 931–936 (2020).
12. Bartolomei, H. et al. Fractional statistics in anyon collisions. *Science* 368,173-177 (2020).
13. Sarma, S., Freedman, M. and Nayak, C. Majorana zero modes and topological quantum computation. *npj Quantum Inf* 1, 15001 (2015).
14. Tsui, D.C., Stormer, H.L. and Gossard, A.C. Two-dimensional magnetotransport in the extreme quantum limit, *Phys. Rev. Lett.* 48, 1559 (1982).
15. Laughlin, R.B. Anomalous quantum Hall effect: An incompressible quantum fluid with fractionally charged excitations. *Phys. Rev. Lett.* 50, 1395 (1983).
16. Willett, R. et al. Observation of an even-denominator quantum number in the fractional quantum Hall effect. *Phys. Rev. Lett.* 59, 1776 (1987).
17. Jain, J.K. Composite-fermion approach for the fractional quantum Hall effect. *Phys. Rev. Lett.* 63, 199 (1989).
18. Jain, J.K. Theory of the fractional quantum Hall effect. *Phys. Rev. B* 41, 7653 (1990).
19. Halperin, B. I. et al. Theory of the half-filled Landau level. *Phys. Rev. B* 47,7312 (1993).
20. Bolotin, K., Ghahari, F., Shulman, M. et al. Observation of the fractional quantum Hall effect in graphene. *Nature* 462, 196–199 (2009).
21. Tang, E., Mei, J.W. & Wen, X.G. High-temperature fractional quantum Hall states. *Phys. Rev. Lett.* 106, 236802 (2011).
22. Tang, H.K. et al. The role of electron-electron interactions in two-dimensional Dirac fermions. *Science* 361,570-574 (2018).
23. Koshino, M., McCann, E. Trigonal warping and Berry's phase $N\pi$ in ABC-stacked multilayer graphene. *Phys. Rev. B* 80, 165409 (2009).
24. Mohammed, M. et al. Topological flat bands in graphene super-moiré lattices. *Phys. Rev. Lett.* 132, 126401 (2024).
25. Abouelkomsan, A. et al. Particle-hole duality, emergent Fermi liquids, and fractional Chern insulators in moiré flatbands. *Phys. Rev. Lett.* 124, 106803 (2020).
26. Chen, G., Sharpe, A.L., Gallagher, P. et al. Signatures of tunable superconductivity in a trilayer graphene moiré superlattice. *Nature* 572, 215–219 (2019).
27. Chen, G., Sharpe, A.L., Fox, E.J. et al. Tunable correlated Chern insulator and ferromagnetism in a moiré superlattice. *Nature* 579, 56–61 (2020).

28. Zhou, H., Xie, T., Taniguchi, T. et al. Superconductivity in rhombohedral trilayer graphene. *Nature* 598, 434–438 (2021).
29. Han, T., Lu, Z., Scuri, G. et al. Orbital multiferroicity in pentalayer rhombohedral graphene. *Nature* 623, 41–47 (2023).
30. Liu, K., Zheng, J., Sha, Y. et al. Spontaneous broken-symmetry insulator and metals in tetralayer rhombohedral graphene. *Nat. Nanotechnol.* 19, 188–195 (2024).
31. Han, T., Lu, Z., Scuri, G. et al. Correlated insulator and Chern insulators in pentalayer rhombohedral-stacked graphene. *Nat. Nanotechnol.* 19, 181–187 (2024).
32. Sha, Y. et al. Observation of a Chern insulator in crystalline ABCA-tetralayer graphene with spin-orbit coupling. *Science* 384, 414-419 (2024).
33. Han, T. et al. Large quantum anomalous Hall effect in spin-orbit proximitized rhombohedral graphene. *Science* 384, 647-651 (2024).
34. Zhou, W., Ding, J., Hua, J. et al. Layer-polarized ferromagnetism in rhombohedral multilayer graphene. *Nat. Commun.* 15, 2597 (2024).
35. Dong, Z., Patri, A. S., and Senthil, T. Theory of fractional quantum anomalous Hall phases in pentalayer rhombohedral graphene moiré structures. *Phys. Rev. Lett.* 133, 206502 (2024).
36. Zhou, B. et al. Fractional quantum anomalous Hall effects in rhombohedral multilayer graphene in the moiréless limit and in coulomb imprinted superlattice. *Phys. Rev. Lett.* 133, 206504 (2024).
37. Dong, J. et al. Anomalous Hall crystals in rhombohedral multilayer graphene I: Interaction-driven Chern bands and fractional quantum Hall states at zero magnetic field. *Phys. Rev. Lett.* 133, 206503 (2024).
38. Guo, Z. et al. Theory of fractional Chern insulator states in pentalayer graphene moiré superlattice. *Phys. Rev. B* 110, 075109 (2023).
39. Herzog-Arbeitman, J. et al. Moiré fractional Chern insulators ii: First-principles calculations and continuum models of rhombohedral graphene superlattices. *Phys. Rev. B* 109, 205122 (2023).
40. Xie, M. & Sarma, S. D. Integer and fractional quantum anomalous Hall effects in pentalayer graphene. *Phys. Rev. B* 109, L241115 (2024).
41. Serlin, M. et al. Intrinsic quantized anomalous Hall effect in a moiré heterostructure. *Science* 367, 900-903 (2020).
42. Park, J.M., Cao, Y., Xia, LQ. et al. Robust superconductivity in magic-angle multilayer graphene family. *Nat. Mater.* 21, 877–883 (2022).
43. Xie, Y., Pierce, A.T., Park, J.M. et al. Fractional Chern insulators in magic-angle twisted bilayer graphene. *Nature* 600, 439–443 (2021).
44. Polshyn, H., Zhu, J., Kumar, M.A. et al. Electrical switching of magnetic order in an orbital Chern insulator. *Nature* 588, 66–70 (2020).
45. Chen, S., He, M., Zhang, YH. et al. Electrically tunable correlated and topological states in twisted monolayer–bilayer graphene. *Nat. Phys.* 17, 374–380 (2021).
46. Grover, S., Bocarsly, M., Uri, A. et al. Chern mosaic and Berry-curvature magnetism in magic-angle graphene. *Nat. Phys.* 18, 885–892 (2022).
47. Tseng, CC., Ma, X., Liu, Z. et al. Anomalous Hall effect at half filling in twisted bilayer graphene. *Nat. Phys.* 18, 1038–1042 (2022).
48. Lin, J.-X. et al. Spin-orbit–driven ferromagnetism at half moiré filling in magic-angle twisted bilayer graphene. *Science* 375, 437–441 (2022).
49. Zhang, Z. et al. Commensurate and incommensurate Chern insulators in magic-angle bilayer graphene. Preprint at https://arxiv.org/abs/2408.12509 (2024).
50. Zhu, J. et al. Voltage-controlled magnetic reversal in orbital Chern insulators. *Phys. Rev. Lett.* 125, 227702 (2020).
51. Shavit. G. & Oreg, Y. Quantum geometry and stabilization of fractional Chern insulators far from the ideal limit. Preprint at https://arxiv.org/abs/2405.09627 (2024).
52. Feng, Z., Wang, W., You, Y. et al. Rapid infrared imaging for rhombohedral graphene. Preprint at https://arxiv.org/abs/2408.09814 (2024).
53. Moon, P. and Koshino, M. Electronic properties of graphene/hexagonal-boron-nitride moiré superlattice. *Phys. Rev. B* 90, 155406 (2014).
54. Zhang, S. H. et al. Correlated insulators, density wave states, and their nonlinear optical response in magic-

angle twisted bilayer graphene. *Phys. Rev. Lett*. 128, 247402 (2022).
55. Miao, W. Q. et al. Truncated atomic plane wave method for subband structure calculations of moiré systems. *Phys. Rev. B* 107, 125112 (2023).

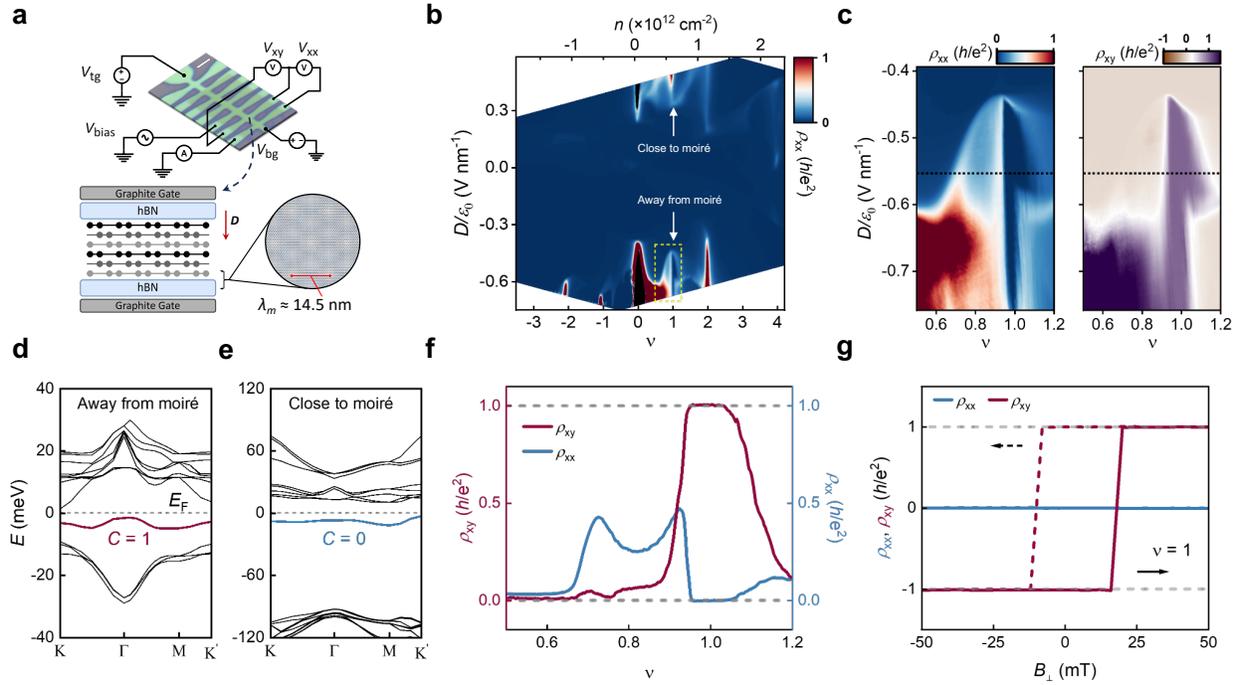

**Fig.1|Topological flat band in RHG/hBN moiré superlattices. a,** Optical image of device 1 and measurement configurations. The scale bar in optical image is 3 μm. The illustration shows the cross-section of the device and a moiré interface between RHG and bottom hBN. **b,** Phase diagram of longitudinal resistivity $\rho_{xx}$ as a function of moiré filling factor ν and electric displacement field $D/\varepsilon_0$ measured at $B_\perp = 0$, demonstrating a series of correlated states at ν = ±1 and ±2. Carrier density $n$ is shown on the top axis. Note that the measurement in the region manually rendered in black is not reliable due to the large resistance and poor contact near the charge neutrality point. **c,** Phase diagrams of longitudinal resistivity $\rho_{xx}$ (left panel) and Hall resistance $\rho_{xy}$ (right panel) measured at the zoom-in region indicated by the yellow dashed box in **b**. **d,** Isolated flat Chern band with $C = 1$ above charge neutrality and right below the chemical potential of ν = 1 in RHG/hBN moiré superlattices calculated at $D/\varepsilon_0 = -0.8$ V nm$^{-1}$ by Hartree-Fock method. **e,** Flat band with $C = 0$ at ν = 1 state calculated at $D/\varepsilon_0 = 0.8$ V nm$^{-1}$, where the electrons are close to the moiré interface. **f,** $\rho_{xx}$ and $\rho_{xy}$ versus ν measured at zero magnetic field and $D/\varepsilon_0 = -0.55$ V nm$^{-1}$ corresponding to the positions indicated by the dashed lines in **c**, featuring quantized $\rho_{xy} = h/\nu e^2$ and vanishing $\rho_{xx}$. **g,** Standard magnetic hysteresis loop of IQAHE at ν = 1 measured at $D/\varepsilon_0 = -0.51$ V nm$^{-1}$. The scanning direction is denoted by different arrows. $\rho_{xy}$ shows sharp transitions at the coercive field $B_c \approx 12$ mT, with a hysteresis window $\Delta B_c \sim 25$ mT, while $\rho_{xx}$ remains vanishing as magnetic field changes.

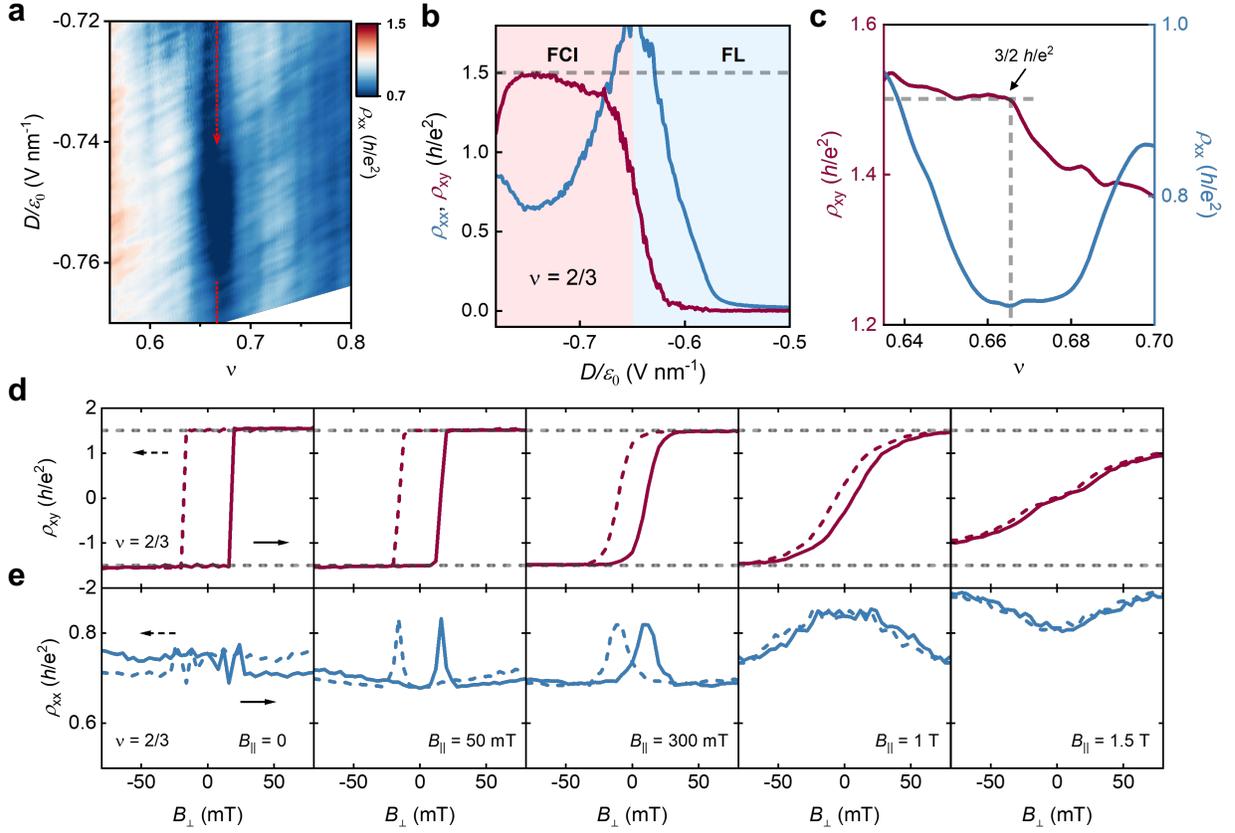

**Fig.2 | The FCI at ν = 2/3. a,** Zoomed-in phase diagram of symmetrized $\rho_{xx}$ measured at $B_\perp = \pm 40$ mT. The red dashed line marks the dip of ν = 2/3. **b,** Symmetrized $\rho_{xx}$ and antisymmetrized $\rho_{xy}$ as a function of electric displacement field $D/\varepsilon_0$, featuring a continuous phase transition between Fermi liquid (FL) and FCI. At an optimal displacement field $D/\varepsilon_0 = -0.75$ V nm$^{-1}$, the $\rho_{xy}$ reaches to a quantized value of $\rho_{xy} = 3h/2e^2$ whereas $\rho_{xx}$ is within a minimum. **c,** Symmetrized $\rho_{xx}$ and antisymmetrized $\rho_{xy}$ as a function of moiré filling factor ν measured at $D/\varepsilon_0 = -0.752$ V nm$^{-1}$. The $\rho_{xx}$ at ν = 2/3 exhibits a clear dip and the $\rho_{xy}$ is approximately quantized to the value of $\rho_{xy} = 3h/2e^2$. **d,e,** Magnetic hysteresis loops of antisymmetrized $\rho_{xy}$ and symmetrized $\rho_{xx}$ measured at different in-plane magnetic fields.

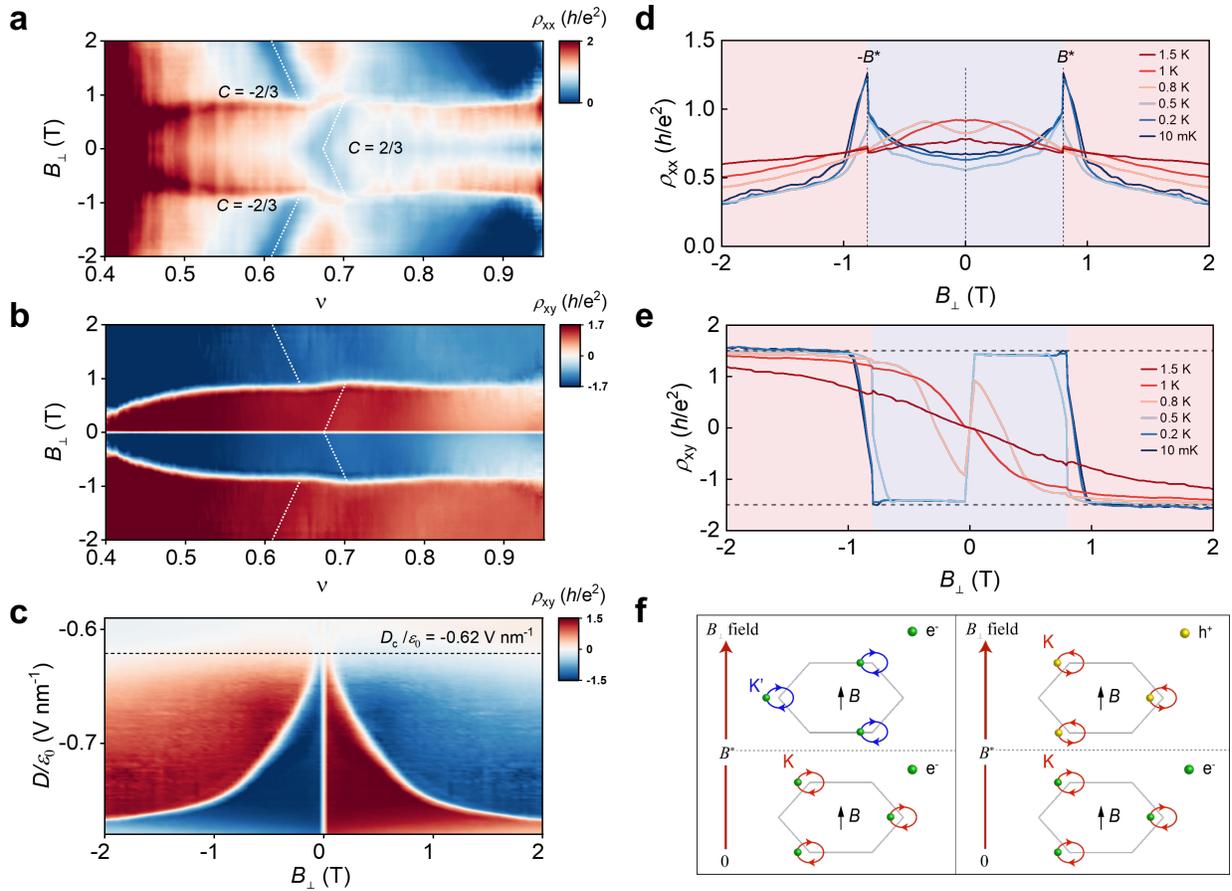

**Fig.3|Phase transitions in the $B_\perp$ field within $0 < \nu < 1$. a,b,** Phase diagrams of symmetrized $\rho_{xx}$ (**a**) and antisymmetrized $\rho_{xy}$ (**b**) as a function of $\nu$ and $B_\perp$ field measured at a displacement field of $D/\varepsilon_0 = -0.74$ V nm$^{-1}$. White dashed lines are guides to the eye for trajectories of $\nu = 2/3$ state with a $|C| = 2/3$ gap. The dashed lines in **b** are assigned based on the position of dips shown in **a**. The trajectories of the Chern gap at $\nu = 2/3$ state are consistent with the behavior of electron-type fermions at low $B_\perp$ fields whereas change to hole-type fermions at high $B_\perp$ fields. **c,** Phase diagrams of antisymmetrized $\rho_{xy}$ as a function of $B_\perp$ field and $D$ field measured at $\nu = 2/3$ state. Below the dashed line corresponding to a critical displacement field $D_c/\varepsilon_0 = -0.62$ V nm$^{-1}$, the $\nu = 2/3$ state starts to evolve into four distinct phases. **d,e,** Symmetrized $\rho_{xx}$ (**d**) and antisymmetrized $\rho_{xy}$ (**e**) versus $B_\perp$ field at different temperatures. The curves at base temperature are extracted along the dashed lines from **a** and **b**, respectively. The curves at higher temperatures are extracted similarly from the data shown in Extended Data Fig. 6. Note that the measurement is carried out slightly below the optimal $D$ field for FCI state to show sizable region of the $C = -2/3$ state in the high $B_\perp$ fields. **f,** Two possible pictures of the quantum phase transitions in the $B_\perp$ field. $B^*$ is the critical $B_\perp$ field to switch the two phases and falls into the curved phase boundaries in **c**. Left panel: The system switches to the K' valley, which hosts a Berry curvature opposite to that of the K valley. Right panel: The two states correspond to the same valley polarization, but charge carrier type transitions from quasi-electron to quasi-hole.

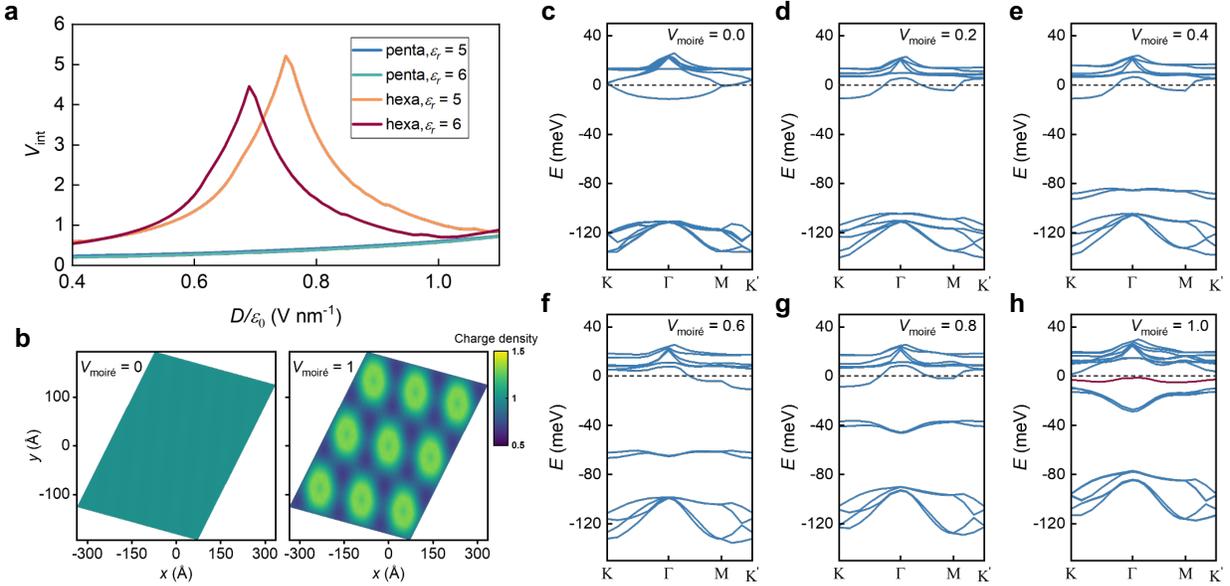

**Fig.4|Calculation of interaction strength, charge density distribution and band structures for hexalayer graphene. a,** Effective interaction strength $V_{int}$ (defined as the ratio between Coulomb interaction energy and the three lowest conduction moiré bands) as a function of $D$ field for pentalayer and hexalayer rhombohedral graphene systems, with background dielectric constants $\varepsilon_r$ = 5 and 6. **b,** Real-space charge density distribution of the Hartree-Fock states at $\nu = 1$ for the hexalayer graphene moiré superlattice. The charge density of the gapless Hartree-Fock state at 0% moiré potential strength is homogeneous everywhere (left panel), with vanishing standard deviation; while the charge density of the Chern insulator state at 100% moiré potential (right panel) exhibits clear charge modulation commensurate with the moiré superlattice. **c-h,** Band structures calculated using cRPA+Hartree-Fock method for RHG system at $\nu = 1$, with different moiré potential strengths. Decreasing the strength of moiré potential would result in gapless metallic state at $\nu = 1$.

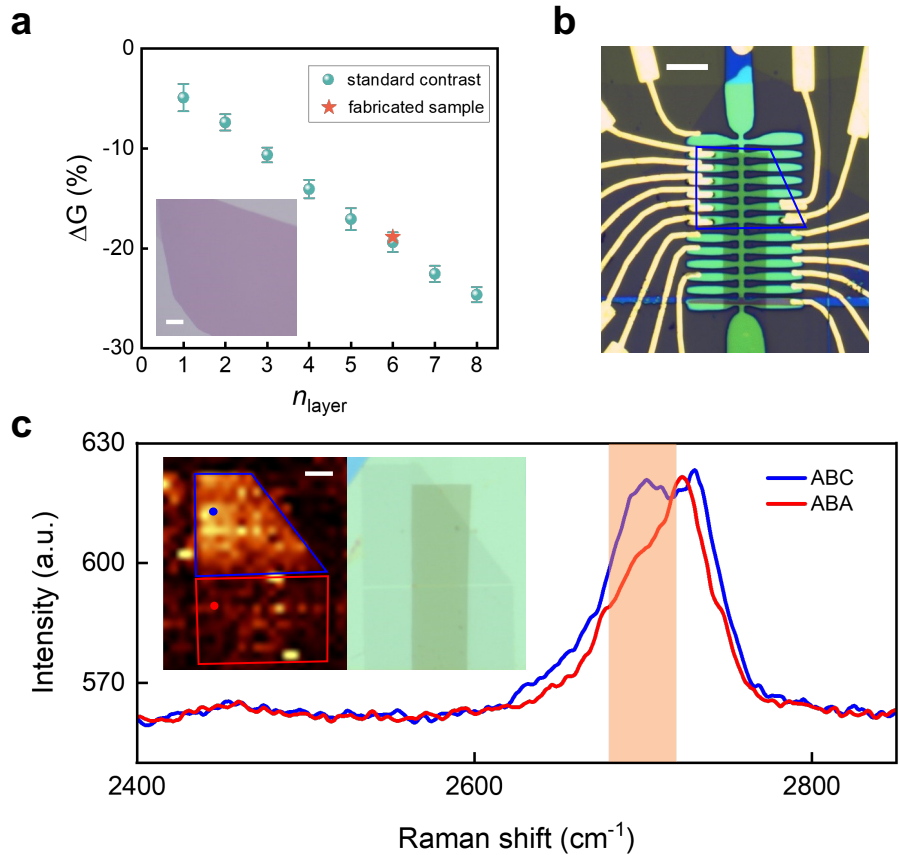

**Extended Data Fig.1 | Identification of layer number and stacking order of the multilayer graphene. a,** Optical contrast parameter ΔG, representing the relative difference of the G values in RGB triplet between calibrated graphene samples and Si substrate, as a function of layer number. The inset shows the optical image of the graphene we fabricated, with a scale bar of 10 μm. The corresponding ΔG value is marked with an orange asterisk. **b,** Final optical image of device 1. The blue box represents the ABC stacking regions. Scale bar is 5 μm. **c,** Raman spectroscopy characterization of the graphene sample after transfer. The blue and red lines show spectroscopy intensity versus Raman shift in ABC stacking and ABA stacking regions, respectively. The integration range is indicated by orange shadow. The inset displays the Raman mapping image (left) and the optical image (right) of the stack, with a scale bar of 5 μm. The blue (top) and red (bottom) boxes represent the ABC and ABA stacking regions, respectively.

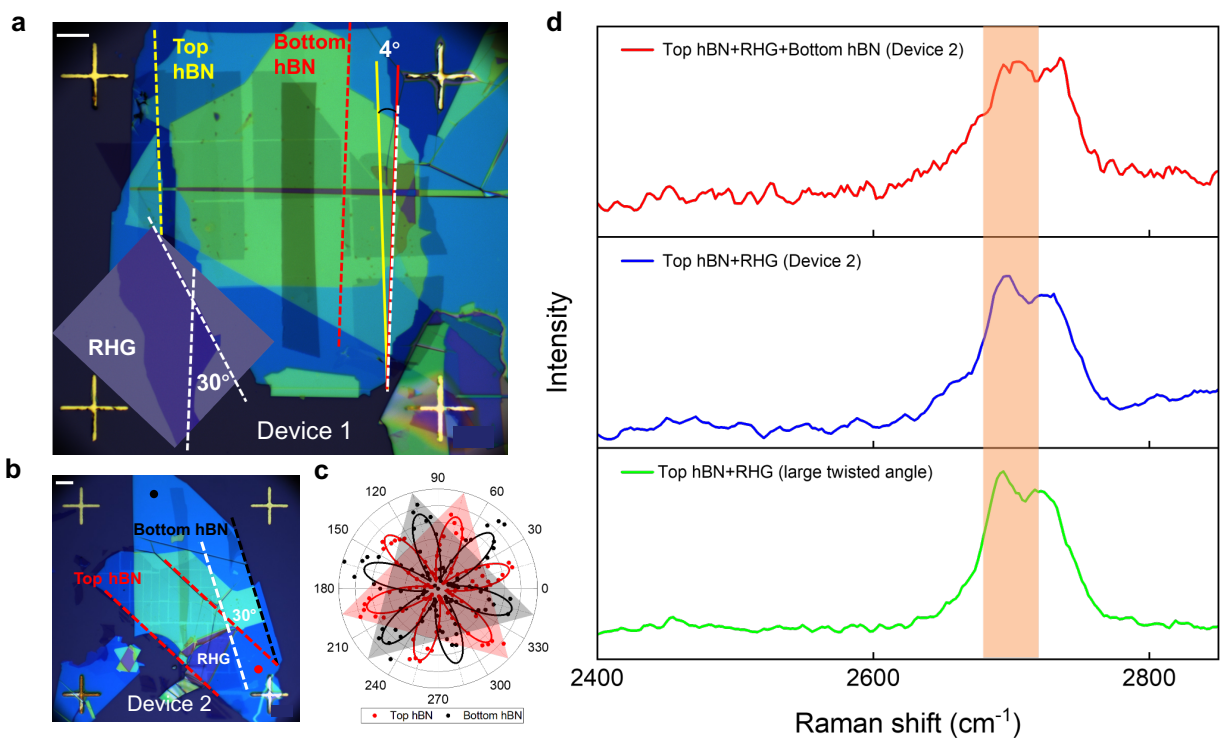

**Extended Data Fig.2|Determination of alignment. a,** Optical image of device 1. The red (yellow) dashed lines indicate the crystallographic edges of bottom (top) hBN. The inset shows the optical image of graphene, where the white dashed lines represent the crystallographic edges. The optical image of graphene has been rotated to align with the position of RHG on device 1. According to the optical images, the twist angle between graphene and top hBN is ~ 4° or ~ 26°. Bottom hBN is well aligned with graphene which is consistent with the large moiré wavelength (~ 14.5 nm, corresponding to a twisted angle of 0.17°) obtained from transport measurements. Scale bar is 10 µm. **b,c,** Determination of relative crystal orientation between top and bottom hBN of device 2. The red (black) dashed lines of the optical image (**b**) represent the crystallographic edges of top (bottom) hBN. Scale bar is 10 µm. Second-harmonic generation (SHG) characterization (**c**) demonstrates that the crystal orientations of top and bottom hBN differ by 30°. Two overlapping triangular patterns in different colors serve as guides to the eye. **d,** Raman spectroscopy characterization of device 2 at different transfer steps, indicating that the RHG in device 2 is aligned with bottom hBN (more details shown in Methods).

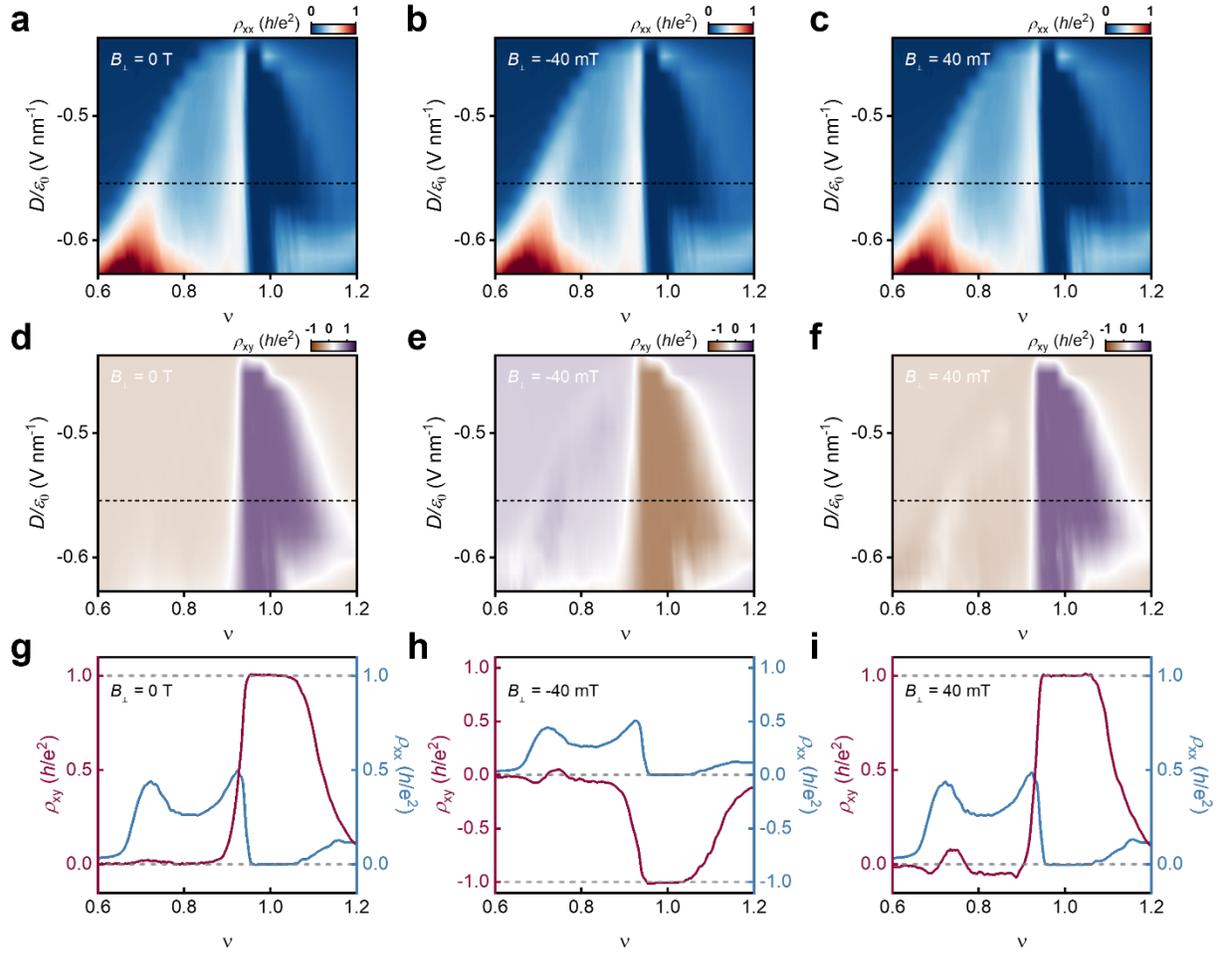

**Extended Data Fig.3|Symmetrized and antisymmetrized processing for $C = 1$ state at ν = 1.**
**a,** Symmetrized $\rho_{xx}$ versus ν and $D/\varepsilon_0$ obtained from symmetrization process of **b** ($B_\perp$ = -40 mT) and **c** ($B_\perp$ = 40 mT). **d,** Antisymmetrized $\rho_{xy}$ versus ν and $D/\varepsilon_0$ obtained from antisymmetrization process of **e** ($B_\perp$ = -40 mT) and **f** ($B_\perp$ = 40 mT). **g,** Line cuts showing vanishing $\rho_{xx}$ and quantized $\rho_{xy}$ taken from **a** and **d**, which are quantitatively consistent with Fig. 1f of the main text. **h,** Line cuts showing vanishing $\rho_{xx}$ and quantized $\rho_{xy}$ taken from **b** and **e**. **i,** Line cuts showing vanishing $\rho_{xx}$ and quantized $\rho_{xy}$ taken from **c** and **f**.

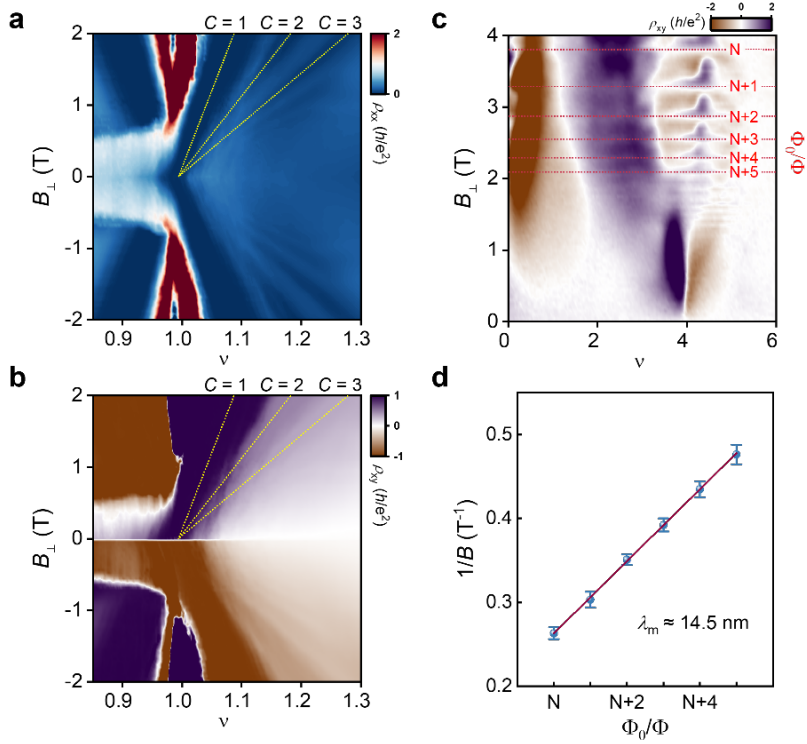

**Extended Data Fig.4 | Calibration of carrier density and twist angle determination. a,b,** Landau fan diagrams of $\rho_{xx}$ (**a**) and $\rho_{xy}$ (**b**) measured at $D/\varepsilon_0$ = -0.71 V nm$^{-1}$ and $T$ = 10 mK (device 1). IQAHE at $\nu$ = 1 features vanishing $\rho_{xx}$ and quantized $\rho_{xy}$. When the magnetic field reaches around 1 T, quantum Hall plateaus with $C$ = 2 and $C$ = 3 appear with yellow dashed lines showing the trajectories. The positions of these states are used to calibrate the carrier density. We can deduce that carrier density of $4n_0$ ($\nu$ = 4) is about 2.18×10$^{12}$ cm$^{-2}$ corresponding to a twist angle $\theta$ ≈ 0.15° with moiré periodicity $\lambda$ ≈ 14.5 nm. **e,f,** Brown-Zak oscillations of device 1 measured at $T$ = 1.5 K (**c**) and linear fitting of the oscillation period (**d**). By fitting 1/$B$ values indicated with red dashed lines, we can obtain a consistent moiré periodicity $\lambda_m$ ≈ 14.5 nm corresponding to a twist angle of $\theta$ ≈ 0.17°.

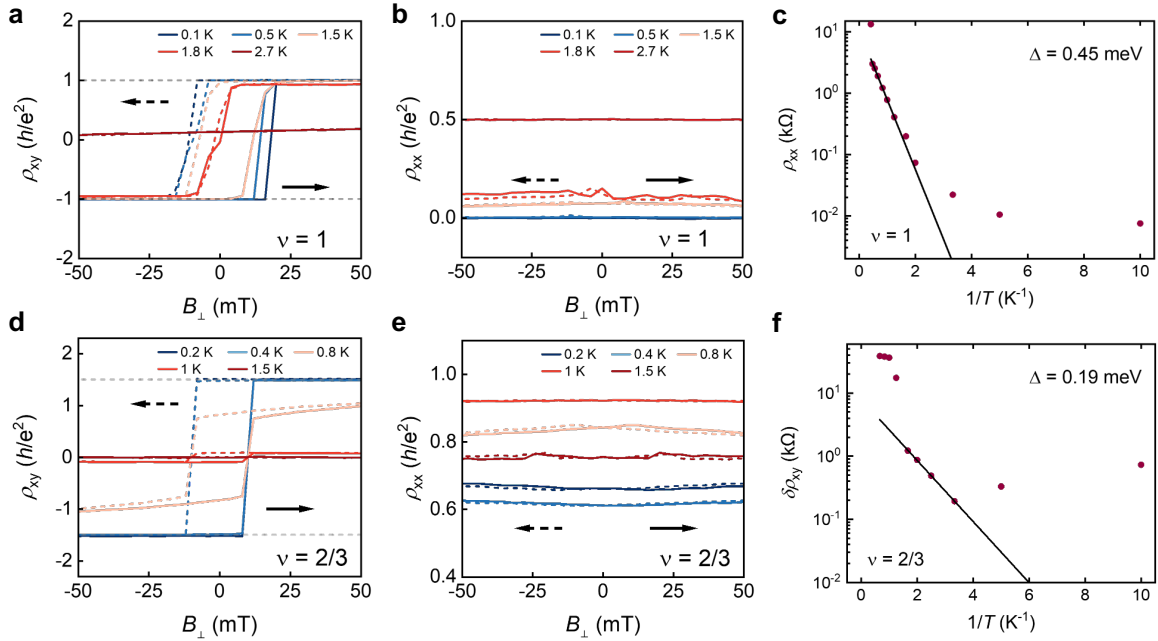

**Extended Data Fig.5 | Temperature dependence and estimation of energy gap at both integer and fractional fillings. a,b,** Temperature dependence of magnetic hysteresis loops at ν = 1. Quantization of $\rho_{xy}$ persists up to 1.5 K. **c,** $\rho_{xx}$ as a function of $1/T$ and the linear fit for ν = 1 state. **d,e,** Temperature dependence of $\rho_{xy}$ and $\rho_{xx}$ scans at ν = 2/3. $\rho_{xy}$ remains quantized at 400 mK and the anomalous Hall signal vanishes when $T > 1$ K. **f,** Plot of $\delta\rho_{xy} = h/(\nu e^2) - \rho_{xy}$ at ν = 2/3. Solid line is the linear fit to $\rho_0 e^{-\Delta/k_B T}$, where $\rho_0$ is the fitting constant and $k_B$ is the Boltzmann constant.

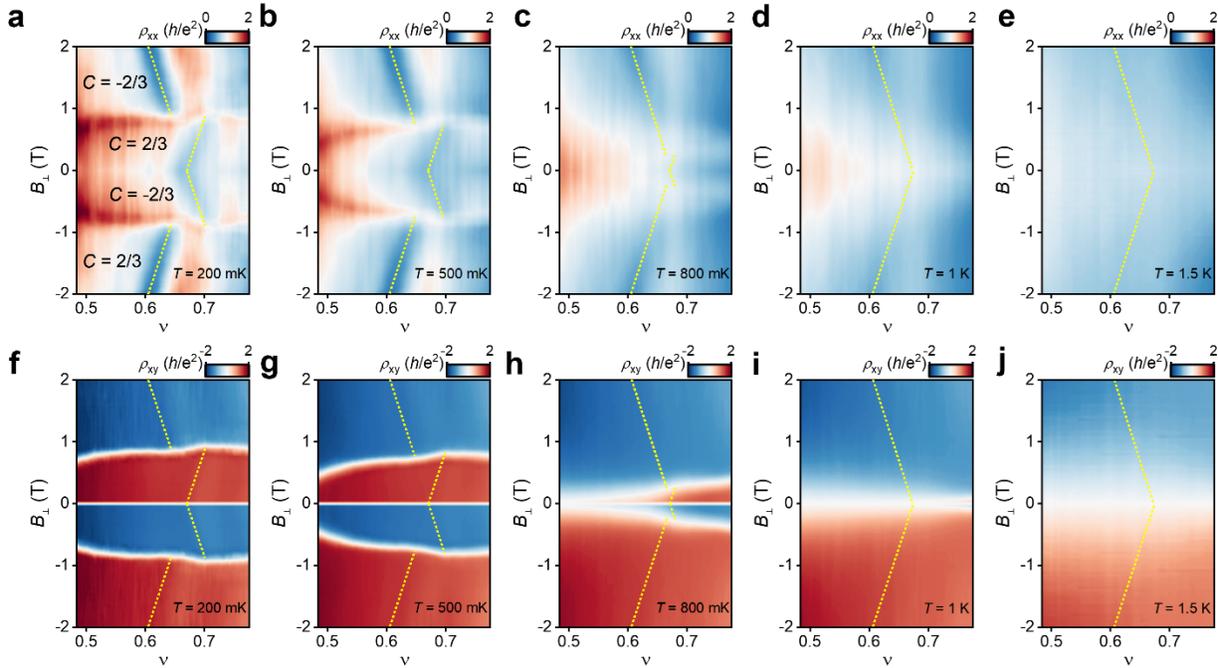

**Extended Data Fig.6|Temperature dependent phase transitions at ν = 2/3.** Phase diagrams of $\rho_{xx}$ (**a-e**) and $\rho_{xy}$ (**f-j**) as a function of ν and $B_\perp$ field measured at $D/\varepsilon_0 = -0.74$ V nm$^{-1}$ and at different temperatures. The yellow dashed lines show the slopes of $C = \pm 2/3$. The dashed lines in $\rho_{xy}$ are assigned based on the position of dips shown in $\rho_{xx}$. Clear phase transition behavior from composite electron to composite hole persists up to 800 mK.

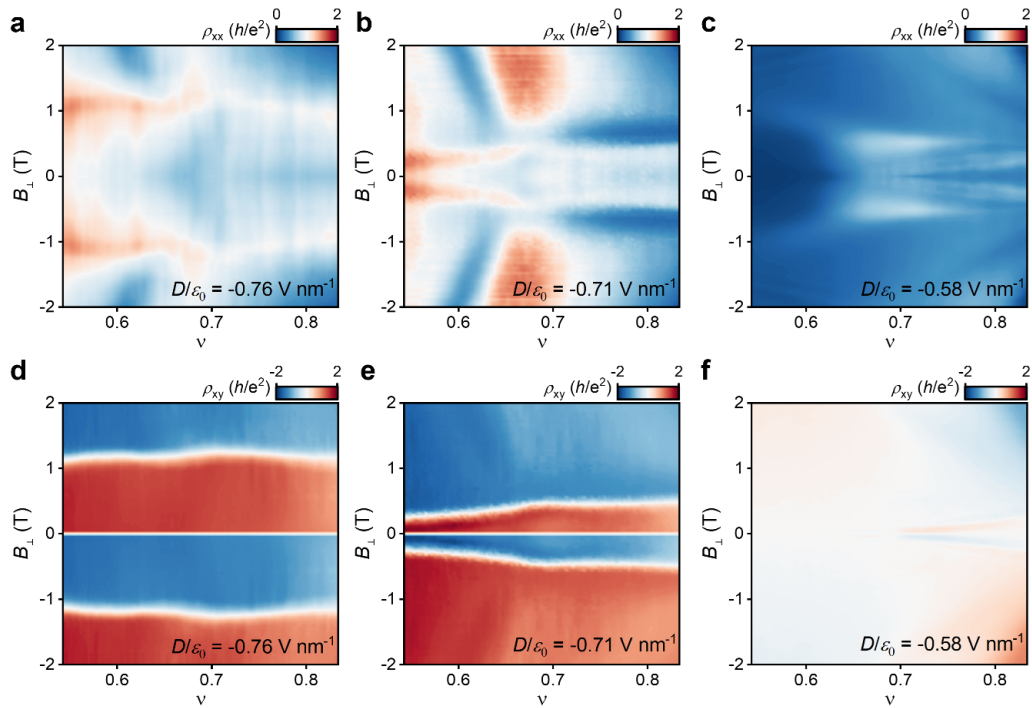

**Extended Data Fig.7|Electrical displacement field tunable phase transitions at ν = 2/3.** Phase diagrams of $\rho_{xx}$ (**a-c**) and $\rho_{xy}$ (**d-f**) as a function of ν and $B_\perp$ field at different $D/\varepsilon_0$. As electrical field decreases (from left to right), the phase boundary gradually shifts towards lower magnetic fields and disappears eventually.

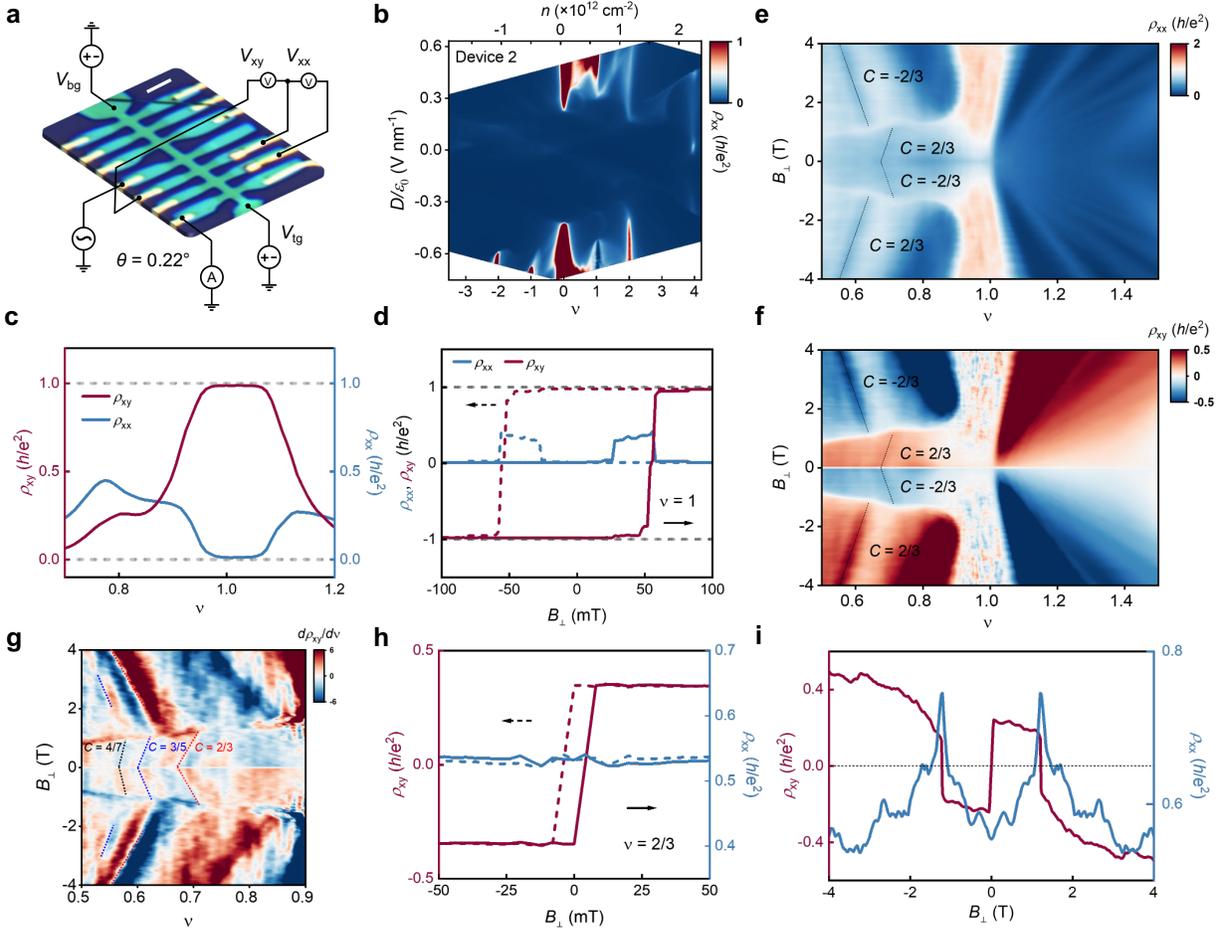

**Extended Data Fig.8 | Topologically nontrivial states observed in device 2. a,** Optical image and schematic of measurement configurations. Scale bar is 3 μm. **b,** Phase diagram of longitudinal resistivity $\rho_{xx}$ as a function of moiré filling factor ν and electric displacement field $D/\varepsilon_0$ measured at $B_\perp = 0$. Carrier density $n$ is shown on the top axis. **c,** Symmetrized $\rho_{xx}$ and antisymmetrized $\rho_{xy}$ as a function of moiré filling factor ν measured at $D/\varepsilon_0 = -0.635$ V nm$^{-1}$, featuring quantized $\rho_{xy} = h/\nu e^2$ and vanishing $\rho_{xx}$. **d,** Magnetic hysteresis scans of ν = 1 state measured at $D/\varepsilon_0 = -0.659$ V nm$^{-1}$. **e,f,** Landau fan diagrams of symmetrized $\rho_{xx}$ (**e**) and antisymmetrized $\rho_{xy}$ (**f**) measured at $D/\varepsilon_0 = -0.833$ V nm$^{-1}$. The 2/3 state features a clear dip in $\rho_{xx}$ and a plateau in $\rho_{xy}$, both of which fully adhere to Streda's formula. The fan diagrams are also used to calibrate the carrier density and the value of $4n_0$ (ν = 4) is about $2.24 \times 10^{12}$ cm$^{-2}$ corresponding to a twist angle of $\theta \approx 0.22°$. **g,** Differential of $\rho_{xy}$ as a function of ν and $B_\perp$ field measured at $D/\varepsilon_0 = -0.833$ V nm$^{-1}$. Dashed lines in different colors serve as visual guides for the trajectories of the ν = 4/7, 3/5 and 2/3 states, respectively. **h,** Magnetic hysteresis scans of ν = 2/3 state measured at $D/\varepsilon_0 = -0.833$ V nm$^{-1}$. **i,** Symmetrized $\rho_{xx}$ and antisymmetrized $\rho_{xy}$ versus $B_\perp$ field. The curves are extracted along the dashed lines from **e** and **f**, respectively.